\documentstyle[12pt,prl,aps,graphicx]{revtex}
\begin{document}
%%%%%%%%%%%%%%%%%%%%
 \title{ Pulsations and stability of stars with phase transition}
\author{\it  Zakir F. Seidov, 
Dept of Physics, Ben-Gurion University\\
\sf E-mail: seidov@bgumail.bgu.ac.il}
\maketitle 
\begin{abstract} 
The general characteristics of stars with first-order phase transition
(PT1) are reviewed at short. The model of 
two-incompressible-phase star with PT1 is considered in some detail.
\end{abstract}
\section{Introduction}
First-order phase transition (PT1) is characterized 
by density jump from $\rho_1$ to
$\rho_2$ with $q=\rho_2/\rho_1>1$, at some pressure $P_0.$ 
PT1 leads to density discontinuity inside the star 
while pressure is continious across boundary between phases.
Also continious are gravity- and pressure-induced 
forces inside the star.
Here we review some results of studying such stars get mainly 
(but {\it not} only) by author.
\section{Three methods}
There are three main methods of star's equilibrium and stability analysis:
\begin{itemize} 
\item Static Criterion (Mass - Central Pressure dependence)
\item Dynamical Principle (pulsation frequency)
\item Variational (Energetic) Principle (variation of total energy).
\end{itemize}

Stability loss/restoration (critical equilibrium
states) according to these methods are defined as follows:
\begin{itemize}
\item Mass extremum, $dM/dP_c=0$
\item Zero frequency of pulsation (of the lowest mode),  $\omega ^2 = 0$
\item Equilibrium energy extremum, $\delta ^2 E = 0.$
\end{itemize}
	In fig.1 two first principles are shown qualitatively. 
Unstable equilibrium states are shown by broken line.
 \begin{figure}
\includegraphics[scale=.5]{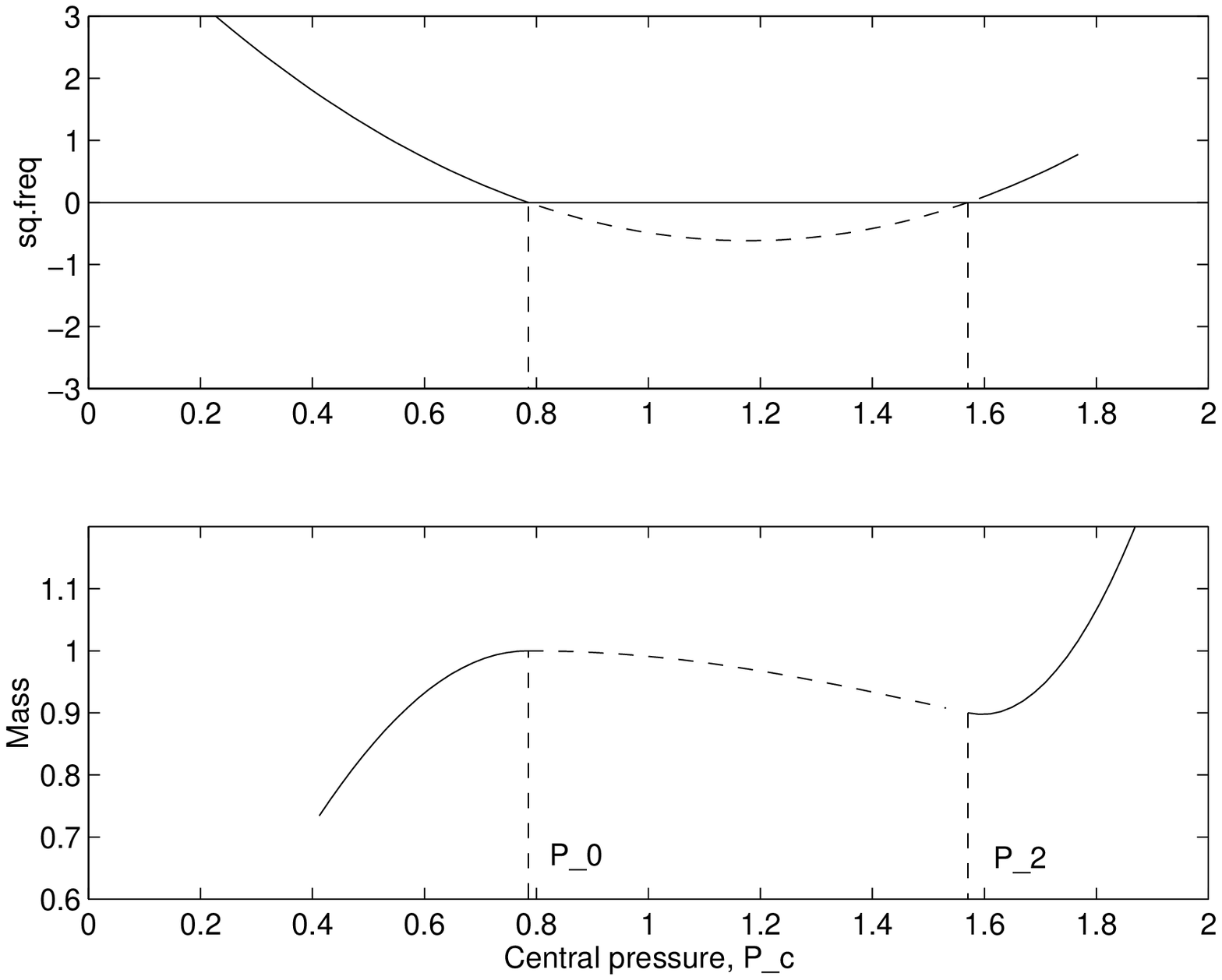}
\caption{Static (lower part) and dynamical (upper part)\\ 
principles in the equilibrium and stability of stars}
 \end{figure} 
\newpage
\section{Newton Theory of Gravitation} 
Here are some general results in NTG for stars with PT1:
\begin{itemize}
\item At $q>1.5$, stability loss occurs at $P_c=P_0$ for 
{\em any} EoS, while
 {\em recover} of stability for larger $P_c$ depends on EoS
\item At $q_{min}(\gamma)<q <1.5$,
 stability loss occurs only at $P_c>P_0$;
{ \sf $\gamma\;=\;1+1/n,\;\gamma \mbox{ and }n$ are adiabatic and
polytropic indices.
E.g., $q_{min}=1.46,\;1.33,\;1.20,\mbox{ and }1.09
\mbox{ for }n=1,\;1.5,\;2\mbox{ and }2.5.$}\\
The critical value of (relative) mass of new-phase core is larger for smaller 
$q$'s and for smaller  $\gamma$'s ($softer$ EoS's).\\
{ \sf E.g., critical value of new-phase core $x_{crit}=8/9\;(q-3/2)
\mbox{ for }\gamma=1\;(n=1)$}
\item
At slow rotation with angular velocity $\Omega$, in
spherical approximation,
$q_{crit}=3/2\;-\;\Omega^2\;/\;4\;\pi\;G\;\rho_1 $, 
for stability loss at $P_c=P_0$ for {\em any} EoS.
\item For PT1, $starting$ at finite radius ("$neutral\;core $"),\\
 the larger size of neutral core, the larger value of $q_{crit}.$\\
{ \sf E.g.,$q_{crit}\rightarrow \infty \mbox{ at } x\rightarrow\;3/4
\mbox{ at } n=0 \mbox{ and } x\rightarrow\;.6824 \mbox{ at } n=1.$}
\end{itemize}
 \subsection{Classical example}
Inverse $\beta$-decay reactions in dense degenerate matter of white dwarf 
stars lead to nuclei transformations 
$(A,Z)\rightarrow (A,Z-2)$ ("$neutronization$"), 
and to density jump with $q=Z/(Z-2)<1.5$.
$M - \rho_c$ curves for equilibrium cold white dwarfs in classical paper by T.Hamada,
E.Salpeter, Ap.J. {\bf {134}} (1961) 669 are $incorrect$, as:\\
a) central density $\rho_c$ is $not$ continious variable, and
b) mass maximum is $not$ at point $P_c=P_0,$ but at some $P_c>P_0.$
\subsection{General Relativity}
In GR, the  critical value of energy density
jump $q=\varepsilon_2/\varepsilon_1$ is equal
to $3/2*(1+P_0/\varepsilon_1)$ 
for stability loss at $P_c=P_0$ for {\em any} EoS.

\section{Two-Incompressible-Phase Star } 
Here are some important features of this model:
\begin{itemize}
\item At $q\;\le\;1.5$ there is no unstable equilibrium states
\item At $q>1.5$ recover of stability occurs (at point $P_c=P_2$) for
relative radius of new-phase core, $x=r_{core}/R_{star}$, defined by
relation:
\begin{equation} f(q,x)=(q-1)^2 x^4 +4\;(q-1)\;x +3 - 2q=0
\end{equation}
{ \sf Star with $x>\surd 2 -1$ is stable for $arbitrary\;large\;q$} 
\item Frequency squared 
of the small adiabatic radial pulsations of the lowest mode 
for star in $slow$ rotation with angular velocity $\Omega$:\\ 
$\omega_\Omega^2=\omega_0 ^2 +\Delta_{\Omega}(q,x);$
\begin{equation}{ \omega_0^2={4\pi G
\rho_1 f(q,x)\over 3(q-1)(1-x)},}\end{equation}
\begin{equation}
\Delta_{\Omega}(q,x)=\frac23 \Omega^2
\left(\frac{5x(1-x)(1+x)^2}{1+(q-1)x^5}-
\frac{1+(q-1)x}{(q-1)(1-x)}\right).\end{equation}
 At small cores ($x\rightarrow 0$), $\Delta_{\Omega}$ is
{\it negative} --
 rotation {\it reduces} the stability of star with PT1. 
 In general, $\Delta_{\Omega}$ may be of both signs, e.g.,
 at $q>2.11$, rotation leads to decreasing
of value of  $x_{crit}$ from Eq.(1).
\end{itemize}
\subsection{Non-linear pulsations}
 In the next approximaion, pulsations are {\it non-harmonic}.\\
Writing down $R=R_{eq}+z,\ \ |z| << R_{eq},\ \ R_{eq}\ $being radius
of equilibrium model with the same mass, we get in next-to-zeroth approximation
the following equation of motion:
%%%%%%%%%%%%%%%%%%%%%%%%%%%
\begin{equation}\ddot {z} +\omega_{0}^2 z +Cz^2+D \dot z^2=0,
\end{equation}
where $ \omega_0^2$ is as in Eq. (2), while constants $C$ and $D$ are 
some functions of $q$ and of equilibrium value of $x$.
The solution of Eq. (4) with accuracy up to $a^2, \;$( $ a\;$ being an
amplitude of z)  is as follows:
\begin{equation}
z(t)=-\frac{1}{2}(\frac{C}{\omega_0^2}+D)a^2+a \times \cos \omega_0 t +
\frac{1}{6}(\frac{C}{\omega_0^2}-D)a^2+a^2\times\cos 2 \omega_0 t.
\end{equation}
In this approximation the frequency does not differ from one in zero'th
approximation, $ \omega_0$, that is period $T=T_0=2\pi / \omega_0$, while
pulsations are non-sinusoidal.\\
An amplitude of a star's expansion is larger than
 an amplitude of a star's compression and star spends more time with
$R>R_{eq}$ than in state with $R<R_{eq}$. 
A character of non-harmonicity - a slow  large-amplitude "expansion" and rapid 
"contraction" with smaller amplitude -  should be the
same for all equilibrium stable models at $\lambda > 3/2$.\\
 {\it Damping} of pulsations of star with PT1 depends largely on
relation between velocity of motion $v_p$ and a sound velocity $v_s$ in
the region of phase transition. \\{\small In general, the larger $v_p/v_s$ the
larger damping effect due to PT.}

\subsection{General Relativity}
Full analytical investigation of equlibrium and stability of 
two-imcompressible-phase model with PT1 is possible
by static and variational methods. \\In first post-Newtonian
approximation, ($P_0 / \varepsilon_1 \ll 1$) critical value of relative "radius" of core,
at which stability recover occurs, \\is equal to:
\begin{equation}
x_{crit}(q)=x+\Delta_{PN}*(P_0/\varepsilon_1),\end{equation}
with $x\equiv x_N$ defined in Eq. (1) and :
\begin{equation}\Delta_{PN}=\frac{9-7q+27(q-1)x+(4q^2-33q+27)x^2+
(q-1)(9-4q)x^3}{2(q-1)(1+(q-1)x^3)^3}\end{equation}
At $q\rightarrow 3/2$, $x_{PN}\rightarrow q-3/2-3/2\;P_0/\varepsilon_1$,
so that $x_{PN}=0$ at $q=3/2\; (1+P_0/\varepsilon_1)$,\\
which coincides with result for $any$ EoS and for $any$, not only for
small values of $P_0/\varepsilon_1$. In general, first post-Newtonian
correction to $x_{crit}$ may be of both signs, negative at $q<1.89$
and positive for larger $q's.$ 
{ \sf At $q=4+\surd 8,\;x=\surd 2-1
+(59-41\surd 2)/2\;P_0/\varepsilon_1 .$}
%%%%%%%%%%%%%%%%%%

In fact, dependence of both corrections, due to GR and rotation, on $x$
and $q$ is rather complicated, and Fig.~2 presents only part of
$(q-x)\--$plane with lines, on which $\Delta_{\Omega}(q,x)=0$ and
$\Delta_{PN}(q,x)=0$. Also shown is the curve $f(q,x)=0$ from Eq. (1) 
which marks a boundary between $stable$ equilibrium states (at right) 
from $unstable$ ones (at left).  
\begin{figure} \includegraphics[scale=.4]{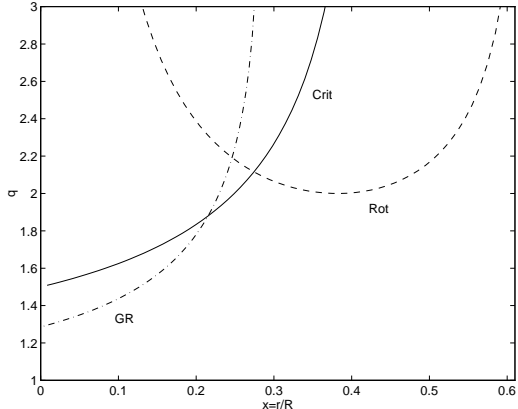}
\caption{Three important curves at $(q-x)\--$plane: critical state of
stability recover, see Eq.~1, and GR and Rotational corrections equal to zero,
see Eq.~(3) and Eq.~(7) respectively.} \end{figure}
\noindent {\large \bf  References}\\
1. W.H. Ramsey, MNRAS {\bf 110} (1950) 325; {\bf 113} (1951) 427;\\
M.J. Lighthill, MNRAS {\bf 110} (1950) 339 ;\\
W.C. De Markus, Astron. J. {\bf 59} (1954) 116 .\\
2. T. Hamada, E.E.Salpeter, Ap.J. {\bf 134} (1961) 669;\\
E. Schatzman, Bull. Acad. Roy. Belgique {\bf 37} (1951) 599;\\
E. Schatzman, White Dwarfs, North-Holland Publ. Co. Amsterdam, 1958.\\
3. Z.F. Seidov, 
Izv. Akad. Nauk Azerb. SSR, ser. fiz-tekh. matem.  no. 5 (1968) 93;\\
Soobsh. Shemakha Astrophys. Observ. {\bf 5} (1970) 58 ;\\
Izv. Akad. Nauk Azerb. SSR, ser. fiz-tekh. matem.  no. 1-2 (1970) 128;\\
Izv. Akad. Nauk Azerb. SSR, ser. fiz-tekh. matem.  no. 6 (1969) 79.\\
4. Ya.B. Zeldovich, Z.F. Seidov (1966) unpublished;\\ 
Z.F. Seidov, Astrofizika {\bf 3} (1967) 189 .\\ 
5. Ya.B. Zeldovich, I.D. Novikov, Relativistic Astrophysics, Univ. Chicago
Press, Chicago, 1971.\\
6. Z.F. Seidov, Astron. Zh. {\bf 48} (1971) 443;\\ 
B. K{\"a}mpfer, Phys.Lett. {\bf 101B} (1981) 366.\\
7. Z.F. Seidov, Astrofizika {\bf 6} (1970) 521.\\
8. Z.F. Seidov, Space Research Inte Preprint (1984) Pr-889.\\
9. M.A. Grienfeld, Doklady Acad. Nauk SSSR {\bf 262} (1982) 134;\\
G.S. Bisnovatyi-Kogan, Z.F. Seidov, Astrofizika {\bf 21} (1984) 570.\\ 
\end{document}